%% file: main.tex
\documentclass[conference,a4paper]{IEEEtran}
\IEEEoverridecommandlockouts

\usepackage{cite}
\usepackage[pdftex]{graphicx}

\usepackage{amsmath}
\usepackage{amssymb}
\usepackage{array}
\usepackage{amsthm}
\newtheorem{theorem}{Theorem}

\newtheorem{definition}{Definition}
\newtheorem{corollary}{Corollary}

\usepackage{siunitx}
\usepackage{tikz}
\usepackage{pgfplots}
\usetikzlibrary{positioning,arrows,patterns}
\pgfplotsset{compat=1.16}
\usepgfplotslibrary{fillbetween}
\usepackage[capitalize]{cleveref}

\newcommand{\R}{\mathbb{R}}
\newcommand{\transpose}{\mathsf{T}}

\newcommand{\grad}{{\boldsymbol\gamma}}

\newcommand{\gradmean}{\zerobf}
\newcommand{\gradcov}{{\boldsymbol\Sigma}}

\newcommand{\ebf}{\mathbf{e}}
\newcommand{\sbf}{\mathbf{s}}
\newcommand{\xbf}{\mathbf{x}}
\newcommand{\Xbf}{\mathbf{X}}
\newcommand{\ybf}{\mathbf{y}}
\newcommand{\zbf}{\mathbf{z}}
\newcommand{\nbf}{\mathbf{n}}
\newcommand{\hbf}{\mathbf{h}}
\newcommand{\gbf}{\mathbf{g}}
\newcommand{\ubf}{\mathbf{u}}
\newcommand{\vbf}{\mathbf{v}}
\newcommand{\Vbf}{\mathbf{V}}
\newcommand{\wbf}{\mathbf{w}}

\newcommand{\abf}{\mathbf{a}}
\newcommand{\Abf}{\mathbf{A}}
\newcommand{\Pbf}{\mathbf{P}}
\newcommand{\Qbf}{\mathbf{Q}}
\newcommand{\Rbf}{\mathbf{R}}

\newcommand{\zerobf}{\boldsymbol{0}}

\newcommand{\Id}{\mathbf{I}}
\newcommand{\Gauss}[2]{\ensuremath{\mathcal{N}\left( {#1}, {#2} \right)}}

\newcommand{\objective}{\tilde{f}}

\DeclareMathOperator{\Exp}{\mathbb{E}}

\DeclareMathOperator{\Cov}{Cov}
\DeclareMathOperator{\diag}{diag}

\DeclareMathOperator{\Trace}{Tr}
\DeclareMathOperator{\snr}{SNR}

\begin{document}
\title{{Secure Over-the-Air Computation using Zero-Forced Artificial Noise}}
\author{
    \IEEEauthorblockN{Luis Maßny, Antonia Wachter-Zeh}%
    \IEEEauthorblockA{School of Computation, Information and Technology, Technical University of Munich (TUM), Munich, Germany\\
    \{\texttt{luis.massny,antonia.wachter-zeh}\}\texttt{@tum.de}}%
    \thanks{
    This work was supported by the Bavarian Ministry of Economic Affairs, Regional Development and Energy within the scope of the 6G Future Lab Bavaria.
    }%
\vspace{-1cm}}

\maketitle

\begin{abstract}
Over-the-air computation has the potential to increase the communication-efficiency of data-dependent distributed wireless systems, but is vulnerable to eavesdropping. We consider over-the-air computation over block-fading additive white Gaussian noise channels in the presence of a passive eavesdropper. The goal is to design a secure over-the-air computation scheme. We propose a scheme that achieves MSE-security against the eavesdropper by employing zero-forced artificial noise, while keeping the distortion at the legitimate receiver small. In contrast to former approaches, the security does not depend on external helper nodes to jam the eavesdropper's received signal. We thoroughly design the system parameters of the scheme, propose an artificial noise design that harnesses unused transmit power for security, and give an explicit construction rule. Our design approach is applicable in both cases, if the eavesdropper's channel coefficients are known and if they are unknown in the signal design. Simulations demonstrate the performance, and show that our noise design outperforms other methods.
\end{abstract}

\section{Introduction}
\label{sec:introduction}
\input{introduction}

\section{Problem Setting}
\label{sec:setting}
\input{setting}

\section{Secure Over-the-Air Computation Scheme}
\label{sec:scheme}
\input{scheme}

\section{System design}
\label{sec:design}
\input{design}

\section{Simulations}
\label{sec:simulations}
\input{simulations}

\bibliographystyle{IEEEtran}  

\appendices
\input{appendices}

\end{document}

%% file: introduction.tex
While in former wireless communication networks, the communication protocols were designed separately from the network services, next generation systems are envisioned to natively integrate a large variety of services~\cite{khan6GWirelessSystems2020}. These services shall be hosted at the wireless or network edge. 
The  massive amounts of data that are exchanged in applications such as Federated Learning (FL)~\cite{mcmahanCommunicationEfficientLearningDeep2017} limit the system performance.

Over-the-air computation has been proposed as an idea to overcome this bottleneck by jointly designing the communication and computation~\cite{nazerComputationMultipleAccessChannels2007, goldenbaumHarnessingInterferenceAnalog2013}. It was shown to be applicable in the context of wireless sensor networks~\cite{abariOvertheairFunctionComputation2016,goldenbaumRobustAnalogFunction2013}, and recently received a lot of attention for the model aggregation in FL over wireless networks~\cite{zhuBroadbandAnalogAggregation2020, yangFederatedLearningOvertheAir2020, seryOvertheAirFederatedLearning2021, amiriFederatedLearningWireless2020}. The main idea is to leverage the superposition property of the linear wireless multiple-access channel and create interference between the users' signals intentionally. Thereby, the computation is implicitly performed over the air. Instead of reconstructing individual user inputs, the receiver extracts the (noisy) computation result directly from the received signal. While opening new opportunities for communication-efficient services, the multiple-access nature of the wireless channel poses challenges on the security of the computation process, as an eavesdropper can overhear the analog communication. This problem, which we refer to as \emph{secure over-the-air computation}, has been previously addressed in~\cite{huSecureTransceiverDesign2022,yanSecurePrivateOvertheAir2022,freySecureOverTheAirComputation2021}. These works achieve security of the computation result through friendly jamming, either by using a full-duplex receiver~\cite{huSecureTransceiverDesign2022}, by adaptively selecting users as jammers~\cite{yanSecurePrivateOvertheAir2022}, or by an additional helper node~\cite{freySecureOverTheAirComputation2021}.
All these previous works consider the mean squared error (MSE) at the eavesdropper as the security metric, which has a strong practical meaning for the achievable computation precision at the eavesdropper. 
In FL applications, the MSE directly relates to convergence properties of the learning algorithm.

The objective of this work is to ensure security of the computation result when employing over-the-air computation over an additive white Gaussian noise (AWGN) multiple-access channel. We consider a set of users, a legitimate receiver, and a passive eavesdropper who tries to guess the computation result from the received signal. Since the computation happens over an analog channel, our goal is to find an
analog transmission scheme that guarantees a large distortion for the computation result at the eavesdropper, while the legitimate receiver can achieve a low distortion. Similar to previous works, we adopt the MSE at the eavesdropper as the security metric. In distinction to these works, however, our scheme does neither rely on an external jamming signal, nor does it require full-duplex transceivers. Our approach is inspired by the idea from~\cite{goelGuaranteeingSecrecyUsing2008, swindlehurstFixedSINRSolutions2009, linSecrecyRateGeneralized2013} to add zero-forced artificial noise to the transmitted signal values. Applying this approach to over-the-air computation, however, requires a sophisticated signal design, since the users have to align the transmitted signal levels. This leads to new trade-offs between security, computation error, and channel quality.

Concurrently and independently from this work, the same idea is used in \cite{liaoOvertheAirFederatedLearning2022} to preserve privacy of the user inputs against an eavesdropper in the context of FL. Additive artificial noise is harnessed for privacy and is designed to minimize its impact on the learning performance at the legitimate receiver. Although we use the same methodology, our focus is different as we do not consider privacy of the input data, but aim at securing the computation result against the eavesdropper.

%% file: setting.tex
\subsection{Over-the-Air Function Computation}
We consider a setting where $M$ mobile users with data $\ubf_1,\dots,\ubf_M \in \R^k$ contribute to a distributed computation task
\begin{equation*}
f(\ubf_1,\dots,\ubf_M)_i=f_i(u_{1,i},\dots,u_{M,i}), \quad i=1,\dots,k
\end{equation*}
with $f_i \colon \R^M \rightarrow \R$. A central server at the base station, who is the legitimate receiver, wants to retrieve the function $f$ and coordinates the process. The functions $f_i$ are nomographic~\cite{buckApproximateComplexityFunctional1979}, i.e. they can be written as
\begin{equation*}
f_i(u_{1,i},\dots,u_{M,i}) = \psi_i \left( \sum_{m=1}^{M} \phi_{m,i} \left( u_{m,i} \right) \right)
\end{equation*}
with pre-processing functions $\phi_{m,i} \colon \R \rightarrow \R$ and post-processing functions $\psi_i \colon \R \rightarrow \R$, cf~\cite{goldenbaumNomographicFunctionsEfficient2015,goldenbaumHarnessingInterferenceAnalog2013}.
W.l.o.g., we consider linear post-processing functions, i.e.
\begin{equation*}
    f_i(u_{1,i},\dots,u_{M,i}) = \sum_{m=1}^{M} \tilde{\phi}_{m,i} \left( u_{m,i} \right)
\end{equation*}
where $\tilde{\phi}_{m,i}, \, m=1,\dots,M$ is the effective pre-processing function at user $m$ for coordinate $i$. We define \mbox{$\grad_m = \left( \tilde{\phi}_{m,1}(u_{m,1}), \dots, \tilde{\phi}_{m,k}(u_{m,k}) \right)^\transpose \in \R^k$} to be the pre-processed\footnote{In the context of distributed machine learning, pre-processing commonly represents a gradient or model computation while the post-processing represents a model update at the central server.}
input data at user $m$ and write the objective function $f$ by means of an effective function $\objective$ as
\begin{equation*}
\sbf = f(\ubf_1,\dots,\ubf_M) = \objective(\grad_1,\dots,\grad_M) = \sum_m \grad_m.
\end{equation*}
For the sake of simplicity, we assume that the inputs are i.i.d. with each following a zero-mean Gaussian distribution, i.e. \mbox{$\grad_m \sim \Gauss{\gradmean}{\gradcov}$}. We also assume that the inputs are normalized to unit power, i.e. $\Exp\left[ \left\| \grad_m \right\|^2 \right] = \Trace\left( \gradcov \right) = 1$.

\subsection{Channel Model}

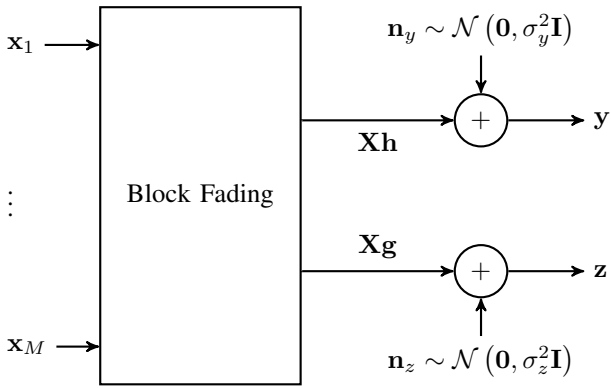
\begin{figure}[!htbp]
    \centering
    \input{channel}
    \caption{AWGN multiple-access wiretap channel with block fading.}
    \label{fig:channel}
\end{figure}
We assume that users, the eavesdropper and the legitimate receiver are single-antenna devices. Furthermore, no communication between the users is allowed, since over-the-air computation is particularly relevant for scenarios where data shall not be exchanged between users, for instance in FL. We model our problem as an AWGN multiple-access wiretap channel with block fading, which is illustrated in \cref{fig:channel}. For transmission, user $m$ maps the input $\grad_m$ onto a transmit signal $\xbf_m \in \R^n$. The channel weights the signal from user $m$ by the block channel coefficient, $h_m$ and $g_m$ respectively, and adds Gaussian noise. This produces output $\ybf$ at the legitimate receiver and $\zbf$ at the eavesdropper. The channel outputs are 
\begin{align*}
    \ybf &= \sum_{m=1}^{M} h_m \xbf_m + \nbf_y = \Xbf \hbf + \nbf_y,\\
    \zbf &= \sum_{m=1}^{M} g_m \xbf_m + \nbf_z = \Xbf \gbf + \nbf_z,
\end{align*}
where $\Xbf = \left( \xbf_1,\dots,\xbf_M \right) \in \R^{n \times M}$. The channel coefficients $\hbf=\left( h_1,\dots,h_M \right)^\transpose \in \R^M$ from the users to the legitimate receiver and $\gbf=\left( g_1,\dots,g_M \right)^\transpose \in \R^M$ from the users to the eavesdropper are independent and distributed according to a Rayleigh distribution with scale $\sigma_h$ and $\sigma_g$ respectively. The channel outputs are distorted by additive Gaussian noise $\nbf_y \sim \Gauss{\zerobf}{\sigma_y^2\Id}$ and $\nbf_z \sim \Gauss{\zerobf}{\sigma_z^2\Id}$ respectively. For the sake of simplicity, we assume real-valued channel coefficients real-valued signals. Our approach and results can be easily generalized to the complex case however.\footnote{If complex signals are considered, the transmission rate increases by a factor of two, since real and imaginary signal parts can be utilized to carry information.}

Perfect channel state information (CSI) for the legitimate receiver is assumed, i.e. the central server knows $\hbf$. We consider both, the case in which the eavesdropper's CSI is unknown, and the case in which it is known to the central server. The latter may be the case in a mobile radio network scenario where the eavesdropper is an active user in the system but untrusted by the legitimate receiver~\cite{huSecureTransceiverDesign2022}.
In addition to the wireless channel, we assume a secure channel from the central server to the trusted contributing users, which is required in order to enable the joint design of the users' transmit signals.

We assume without loss of generality that $h_1 \leq \dots \leq h_M$.
The transmit power at each user is limited by the power constraint \mbox{$\Exp\left[ \left\| \xbf_m \right\|^2 \right] \leq P$}. As it is common for over-the-air computation~\cite{gastparUncodedTransmissionExactly2008}, we consider an uncoded transmission in the sense that $\xbf_m$ is given by precoding of $\grad_m$, but no channel code is applied, and thus, it holds $k=n$. Thereby, it is guaranteed that the receiver can decode (estimate) the objective $\sbf$ from the received value $\ybf$. 
The uncoded transmission is motivated by~\cite{gastparUncodedTransmissionExactly2008}, which shows that uncoded transmission achieves the minimum distortion at the legitimate receiver in the given setting. Formally, we define the distortion by means of the MSE.
As a security measure, we adopt the concept of MSE-security from~\cite{freySecureOverTheAirComputation2021}.
\begin{definition}
Let $\objective \colon \R^k \rightarrow \R^k$ be the objective function. Let $D,S \geq 0$  be real numbers. We say that an over-the-air computation scheme is MSE-approximate with approximation level $D$ if there exists an estimator $d_y \colon \R^n \rightarrow \R^k$, such that
\begin{equation*}
    \Exp\left[ \left\| d_y(\ybf) - \objective\left( \grad_1,\dots,\grad_M \right) \right\|^2 \right] \leq D.
\end{equation*}
We say that an over-the-air computation scheme is MSE-secure with security level $S$ if for every estimator $d_z \colon \R^n \rightarrow \R^k$, we have
\begin{equation*}
    \Exp\left[ \left\| d_z(\zbf) - \objective\left( \grad_1,\dots,\grad_M \right) \right\|^2 \right] \geq S.
\end{equation*}
The expectation is taken over the joint distribution of $\left(\grad_1,\dots,\grad_M\right)$ and $\ybf$ or $\zbf$ respectively.
\end{definition}

%% file: channel.tex
\begin{tikzpicture}
\tikzset{
    >=stealth',
    block/.style={thick, draw, rectangle, inner sep=1em},
    connection/.style={thick,->}
}

\node (tx1) {$\xbf_1$};
\node[below=2cm of tx1.west, anchor=west] (dots) {$\vdots$};
\node[below=2cm of dots.west, anchor=west] (txM) {$\xbf_M$};
\node[block, right=1cm of dots, minimum height=5cm] (macwt) {Block Fading};
\coordinate[above left=2cm and 0cm of macwt.west] (input1);
\coordinate[below left=2cm and 0cm of macwt.west] (inputM);
\coordinate[above right=1cm and 0cm of macwt.east] (top_branch);
\coordinate[below right=1cm and 0cm of macwt.east] (bottom_branch);
\node[draw, circle, thick, right=2cm of top_branch] (main_channel) {$+$};
\node[draw, circle, thick, right=2cm of bottom_branch] (wt_channel) {$+$};
\node[right=1cm of main_channel] (rx) {$\ybf$};
\node[right=1cm of wt_channel] (eve) {$\zbf$};

\node[above=0.5cm of main_channel] (u) {$\nbf_y \sim \Gauss{\zerobf}{\sigma_y^2\Id}$};
\node[below=0.5cm of wt_channel] (r) {$\nbf_z \sim \Gauss{\zerobf}{\sigma_z^2\Id}$};

\draw[connection] (tx1) -- (input1);
\draw[connection] (txM) -- (inputM);

\draw[connection] (macwt) -- (top_branch) --node[below=0.1em] {$\Xbf \hbf$} (main_channel);
\draw[connection] (main_channel) -- (rx);
\draw[connection] (macwt) -- (bottom_branch) --node[above=0.1em] {$\Xbf \gbf$} (wt_channel);
\draw[connection] (wt_channel) -- (eve);
\draw[connection] (u) -- (main_channel);
\draw[connection] (r) -- (wt_channel);

\coordinate[above left=1cm of tx1] (a);
\coordinate[below left=1cm of txM] (b);
\coordinate[right=5cm of b] (c);
\coordinate[right=5cm of a] (d);

\coordinate[below left=0.5cm and 2cm of r.south] (e);
\coordinate[right=4cm of e] (f);
\coordinate[above=3cm of f] (g);
\coordinate[above=3cm of e] (h);

\coordinate[above left=0.5cm and 2cm of u.north] (i);
\coordinate[right=4cm of i] (j);
\coordinate[below=3cm of j] (k);
\coordinate[below=3cm of i] (l);

\end{tikzpicture}

%% file: scheme.tex
The core idea of our secure over-the-air computation scheme is to employ carefully designed additive artificial noise such that it confuses the eavesdropper while not impacting the legitimate receiver.

\subsection{General framework}
The transmit signal for user $m$ is defined as
\begin{equation*}
    \xbf_m = c h_m^{-1} \grad_m + \wbf_m.
\end{equation*}
For each user, we scale the input by the inverse of the channel coefficient in order to compensate for the fading. Next, the signal is scaled by a factor $c$ to meet the power constraint. Note that a common scaling factor has to be used by all users in order to guarantee an unbiased result at the legitimate receiver.
Finally, we add artificial noise by a random vector $\wbf_m \in \R^n$. We take on a zero-forcing approach for the artificial noise design. The artificial noise signals are jointly designed as
\begin{equation*}
    \left( \wbf_1,\dots,\wbf_M \right) = \Vbf \Abf
\end{equation*}
where $\Abf \in \R^{M-1 \times M}$ and the rows of $\Abf$ are a basis of the null space of the channel vector $\hbf$, i.e. $\Abf \hbf = \zerobf$. Generating the artificial noise within the null space of the legitimate receiver's channel makes sure that it is zero-forced for the legitimate receiver and only affects the eavesdropper. We choose the entries of the matrix $\Vbf \in \R^{n \times M-1}$ i.i.d according to $\Gauss{0}{1}$. The artificial noise signals can be jointly designed at the legitimate receiver, who then feeds back the values over a secure channel to the users. Therefore, CSI is only required at the base station. The particular design of the signal scaling $c$ and precoding matrix $\Abf$ is discussed in \cref{sec:design}. Let $\abf_m$ denote the $m$-th column of $\Abf$. Then $\Exp\left[ \left\| \wbf_m \right\|^2 \right] = \left\| \abf_m \right\|^2$. The signal power for user $m$ is thus, given as
\begin{equation}
    \label{eq:signal_power}
    \Exp\left[ \left\| \xbf_m \right\|^2 \right] = \frac{c^2}{h_m^2} + \left\| \abf_m \right\|^2.
\end{equation}
Together with the power constraint this yields an upper bound on the signal scaling factor
\begin{equation}
\label{eq:c}
c^2 \leq \min_m \left\{ h_m^2 \left( P - \left\| \abf_m \right\|^2 \right) \right\}.
\end{equation}

By the definition of the transmitted user signals, the received signals at the legitimate receiver and the eavesdropper are
\begin{align*}
    \ybf &= c \sum_{m=1}^{M} \grad_m + \nbf_y = c \sbf + \nbf_y,\\
    \zbf &= c \sum_{m=1}^{M} \frac{g_m}{h_m} \grad_m + \Vbf \Abf \gbf + \nbf_z = c \sbf + \nbf_e,
\end{align*}
where $\nbf_e = c \sum_{m=1}^{M} \left( \frac{g_m}{h_m}-1 \right) \grad_m + \Vbf \Abf \gbf + \nbf_z$ is the effective noise at the eavesdropper. Note that $\nbf_e$ is a zero-mean Gaussian random vector with covariance
\begin{equation*}
    \Cov\left[ \nbf_e \right] = c^2 \sum_{m=1}^{M} \left( \frac{g_m}{h_m}-1 \right)^2 \gradcov + \left\| \Abf \gbf \right\|^2 + \sigma_z^2 \Id
\end{equation*}
and $\nbf_e$ is correlated with $\sbf$.

The achievable signal-to-noise ratio (SNR) at the legitimate receiver is given as
\begin{equation}
\label{eq:snr}
    \snr = \frac{c^2 M}{\sigma_y^2} \leq \frac{h_1^2 P M}{\sigma_y^2}
\end{equation}
where the upper bound can be achieved only if the artificial noise power vanishes for the weakest user, i.e. $\Exp\left[ \left\| \wbf_1 \right\|^2 \right] = 0$. 

\subsection{Security and approximation}

\begin{theorem}
\label{thm:DvsS}
Consider the proposed over-the-air computation scheme with inputs $\grad_1,\dots,\grad_M$ and objective function $\objective$. For outputs $\ybf$ and $\zbf$ that are jointly Gaussian with $\objective(\grad_1,\dots,\grad_M)$, the computation result is MSE-approximate with approximation level
\begin{equation}
     \label{eq:D-approximate}
     D = M - c^2 M^2 \Trace\left( \gradcov \left( c^2 M \gradcov + \sigma_y^2 \Id \right)^{-1} \gradcov \right),
\end{equation}
and MSE-secure with security level
\begin{equation}
     \label{eq:S-secure}
     S = M - c^2 \left( \sum_{m=1}^{M} \frac{g_m}{h_m} \right)^2 \Trace\left( \gradcov \Rbf^{-1} \gradcov \right)
\end{equation}
and
\(
\Rbf = c^2 \left( \sum_{m=1}^{M} \frac{g_m^2}{h_m^2} \right) \gradcov + \left( \left\| \Abf \gbf \right\|^2 + \sigma_z^2 \right) \Id
\).
\end{theorem}
\begin{IEEEproof}
The statement follows from the covariances of the respective minimum MSE estimators~\cite[p. 155]{poorIntroductionSignalDetection1998}. We refer to Appendix~\ref{apx:DvsS} for the detailed proof.
\end{IEEEproof}

For the special case of i.i.d. coordinates of the users' inputs, we can simplify the results as follows.
\begin{corollary}
If the input coordinates are i.i.d, i.e. $\gradcov = \frac{1}{k} \Id$, the proposed scheme is MSE-approximate with approximation level
\begin{equation*}
     D =
     M - \frac{c^2 M^2}{c^2 M + k \sigma_y^2},
\end{equation*}
and MSE-secure with security level
\begin{equation*}
     S =
     M - \frac{c^2 \left( \sum_{m=1}^{M} \frac{g_m}{h_m} \right)^2}{ c^2 \sum_{m=1}^{M} \frac{g_m^2}{h_m^2} + k \left\| \Abf \gbf \right\|^2 + k \sigma_z^2}.
\end{equation*}
\end{corollary}
By applying the Cauchy-Schwarz inequality
\begin{align*}
    \left( \sum_{m=1}^{M} \frac{g_m}{h_m} \right)^2 \leq M \sum_{m=1}^{M} \frac{g_m^2}{h_m^2}
\end{align*}
it can be observed that the security level given by \eqref{eq:S-secure} relies on two variables: first, the mismatch between the channel coefficients~$\sum_{m} \frac{g_m^2}{h_m^2}$; second, the additional artificial noise contribution~$\left\| \Abf \gbf \right\|^2$. While the former is fixed by the channel realization, the latter can be designed by the choice of the precoding matrix $\Abf$. To satisfy the power constraint, the precoding matrix must be designed with respect to $c$. Therefore, the parameters have to be chosen carefully, see \cref{sec:design}.

The results show that there is a general trade-off between approximation of the computation result at the legitimate receiver and security against the eavesdropper. While the approximation error at the legitimate receiver improves by increasing the signal scaling factor $c$, the security against the eavesdropper decreases at the same time. It can be observed from \eqref{eq:c} that the magnitude of $c$ and thus, the approximation error at the legitimate receiver is lower-bounded by the weakest user's channel and the artificial noise power.

%% file: design.tex
The secure over-the-air computation framework presented in \cref{sec:scheme} can be tailored to the individual requirements by the design of two main variables. First, we can design the MSE at the legitimate receiver based on a Quality of Service (QoS) requirement by adjusting the signal scaling factor $c$. Second, the design of the precoding matrix $\Abf$ is crucial for the achievable security against an eavesdropper. It is possible to either design for approximation at the legitimate receiver first, or design for security against the eavesdropper first. In the following, we consider the former approach, since a practical system needs to set the legitimate receiver's QoS first, and we explore the different design options.

\subsection{Design of the signal scaling}
\begin{figure}[!htbp]
    \centering
    \input{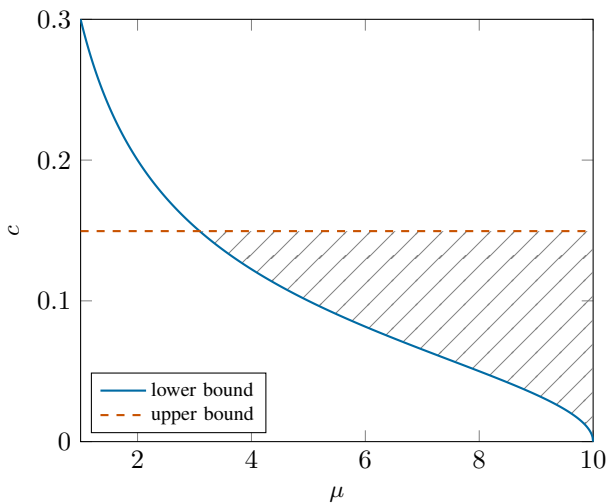}
    \caption{Bounds for the signal scaling coefficient $c$ over the MSE requirement $\mu$ for $M=10$, $k=1$, $\lambda_1=\frac{1}{k}$, and $h_1=\frac{1}{4} \left( \sqrt{\pi} - \sqrt{4-\pi} \right) / \sqrt{2}$, $P = 1$, $\sigma_y^2=0.1$. Highlighted is the achievable design space.}
    \label{fig:c_over_beta}
\end{figure}

Based on \cref{thm:DvsS} and taking into account \eqref{eq:c}, we can design the signal scaling factor $c$ from the Qos requirement for the legitimate receiver. In the following theorem, we state the possible design options. The result is visualized in \cref{fig:c_over_beta}.

\begin{theorem}
\label{thm:mse_requirement}
For a minimum MSE requirement $\mu \leq M$ at the legitimate receiver, the signal scaling factor $c$ must satisfy
\begin{equation}
\label{eq:scaling_by_MSE}
\frac{\sigma_y^2 (M-\mu)}{M^2 k \lambda_1^2 - M(M-\mu) \lambda_1}
\leq c^2 \leq
h_1^2 P
\end{equation}
where $\lambda_1$ is the largest eigenvalue of $\gradcov$.
\end{theorem}
\begin{IEEEproof}
We refer to Appendix~\ref{apx:mse_requirement} for the proof.
\end{IEEEproof}

Note that we can improve the lower bound in \cref{thm:mse_requirement} by considering a more general bound than in \cref{eq:sum_bound} and \cref{eq:ev_bound} as follows. It holds~\cite[Eq. 2]{fultonEigenvaluesInvariantFactors2000} that $\kappa_l \geq \frac{c^2M}{\lambda_i} + \frac{\sigma_y^2}{\lambda_j^2}$ for $i+j=l+n$. In order to choose the tightest bound, we use $i=l$ and $j=n$, and thus, we can bound the eigenvalues as
\begin{equation*}
\kappa_l \geq \frac{c^2M}{\lambda_l} + \frac{\sigma_y^2}{\lambda_n^2}.
\end{equation*}
In total, we obtain
\begin{equation*}
M-\mu \leq c^2 M^2 \sum_{i=1}^{n} \frac{1}{ \frac{c^2M}{\lambda_i} + \frac{\sigma_y^2}{\lambda_n^2} } = M g(c^2)
\end{equation*}
where $g \colon \R_{>0} \rightarrow \R, x \mapsto \sum_{i=1}^{n} \frac{Mx}{ \frac{M}{\lambda_i} x + \frac{\sigma_y^2}{\lambda_n^2} }$ is a strictly monotonically increasing function. Hence, its inverse exists and we can write
\begin{equation*}
c^2 \geq g^{-1}\left( 1 - \frac{\mu}{M} \right)
\end{equation*}
This, however, does not yield a closed-form solution. 
Also note that the lower bound in \cref{thm:mse_requirement} is exact for diagonal covariances $\gradcov$ in particular.

From the upper bound in \eqref{eq:scaling_by_MSE} and together with~\eqref{eq:D-approximate}, we can see that the feasible $\mu$ are restricted to
\begin{equation*}
    \mu \geq M - h_1^2 P M^2 \Trace\left( \gradcov \left( h_1^2 P M \gradcov + \sigma_y^2 \Id \right)^{-1} \gradcov \right).
\end{equation*}
In the system design, we choose the signal scaling factor $c$ according to the lower bound from \cref{thm:mse_requirement}. Therefore, we can rewrite the results from \cref{thm:DvsS}, and obtain the MSE-approximation level and MSE-security level of the proposed secure over-the-air computation scheme for a given MSE requirement $\mu$.

\subsection{Design of the noise precoding}
The precoding matrix $\Abf$ is one of the key aspects in our approach. Simply obtaining a basis of the null space of the channel vector $\hbf$ can be achieved by the SVD as described in~\cite{goelGuaranteeingSecrecyUsing2008}. A distinguishing property of over-the-air computation, however, is that all users have to align the transmit signal levels. Therefore, all users have to reduce the power for the information signal part according to the weakest user. This leads to a potentially great amount of unexploited transmit power, which can be harnessed for security by investing it into the artificial noise. The MSE at the eavesdropper is then maximized by maximizing $\left\| \Abf \gbf \right\|^2$. This is a direct consequence from \cref{thm:DvsS}.

In the following, we carefully design a basis of the null space and optimize it in order to maximize the eavesdropper's received noise power. We start by defining a basis in reduced row echelon form $\Abf^\prime = \left( \abf^\prime_1,\dots,\abf^\prime_M \right) \in \R^{M-1 \times M}$ such that
\begin{equation*}
\abf^\prime_m =
\begin{cases}
\left( \frac{-h_1}{h_M},\dots,\frac{-h_{M-1}}{h_M} \right)^\transpose & \text{if } m=M \\
\ebf_i & \text{otherwise}
\end{cases}
.
\end{equation*}
In order to optimize the power allocation, we introduce scaling coefficients $d_1,\dots,d_{M-1}$ for each row of $\Abf^\prime$. Note that by scaling of the rows in $\Abf^\prime$ we can implicitly scale the columns $\abf_m^\prime$, while preserving the zero-forcing property. The desired basis is obtained as
\begin{equation*}
\Abf = \diag\left(d_1,\dots,d_{M-1}\right) \Abf^\prime .
\end{equation*}
By design it then holds
\begin{equation*}
\left\| \abf_m \right\|^2 =
\begin{cases}
\sum_{i=1}^{M-1} d_i^2 \frac{h_i^2}{h_M^2} & \text{if } m=M \\
d_m^2 & \text{otherwise}
\end{cases}
.
\end{equation*}
We determine the optimal scaling coefficients by a linear optimization program. From the power constraint and \eqref{eq:signal_power} we can obtain the total power budget for artificial noise at user $m$ as
\begin{equation}
\label{eq:power_budget}
\left\| \abf_m \right\|^2 \leq P - \frac{c^2}{h_m^2},
\end{equation}
from which we can derive the conditions in \eqref{condition:budget_other} and \eqref{condition:budget_M}.

We first consider that the eavesdropper's CSI is not known and design the precoding matrix such that it maximizes
\begin{equation*}
    \Exp\left[ \left\| \Abf \gbf \right\|^2 \right] = \sigma_g^2 \sum_{m=1}^{M-1} d_m^2 \left( 2\frac{h_m^2}{h_M^2} - \pi\frac{h_m}{h_M} + 2 \right)
    .
\end{equation*}
In summary, we can formulate a linear optimization program (in $d_m^2$) as follows:
\begin{subequations}
\begin{alignat}{2}
& \!\max_{d_1^2,\dots,d_{M-1}^2 \in \R} &\quad& \sum_{m=1}^{M-1} d_m^2 \left( \frac{h_m^2}{h_M^2} - \frac{\pi}{2}\frac{h_m}{h_M} + 1 \right)\\
& \text{subject to} & & d_m^2 \leq P - \frac{c^2}{h_m^2},\label{condition:budget_other}\\
& & & \sum_{m=1}^{M-1} d_m^2 \frac{h_m^2}{h_M^2} \leq P - \frac{c^2}{h_M^2},\label{condition:budget_M}\\
& & & d_m^2 \geq 0.
\end{alignat}
\end{subequations}

\subsection{Leveraging channel information about the eavesdropper}

If the central server has access to the eavesdropper's CSI, then we can adapt the optimization problem by taking into account the eavesdropper's channel coefficients as follows. We explicitly calculate
\begin{align*}
\left\| \Abf \gbf \right\|^2 &= \sum_{i,j=1}^{M} g_i g_j \abf_i^\transpose \abf_j\\
&=
\sum_{i,j=1}^{M-1} g_i g_j \abf_i^\transpose \abf_j
+ 2 \sum_{i=1}^{M-1} g_i g_M \abf_i^\transpose \abf_M
+ g_M^2 \left\| \abf_M \right\|^2\\
&= \sum_{i=1}^{M-1} d_i^2 \left( \frac{h_i^2}{h_M^2} g_M^2 - 2 \frac{h_i}{h_M} g_i g_M + g_i^2 \right)
\end{align*}
and modify the optimization subject into
\begin{subequations}
\begin{alignat}{2}
& \!\max_{d_1^2,\dots,d_{M-1}^2 \in \R} &\quad& \sum_{m=1}^{M-1} d_m^2
\left( \frac{h_m^2}{h_M^2} g_M^2 - 2 \frac{h_m}{h_M} g_m g_M + g_m^2 \right) \label{eq:optProb}\\
& \text{subject to} & & d_m^2 \leq P - \frac{c^2}{h_m^2}, \label{eq:constraint1}\\
& & & \sum_{m=1}^{M-1} d_m^2 \frac{h_m^2}{h_M^2} \leq P - \frac{c^2}{h_M^2},\label{eq:constraint2}\\
& & & d_m^2 \geq 0.\label{eq:constraint3}
\end{alignat}
\end{subequations}

%% file: simulations.tex
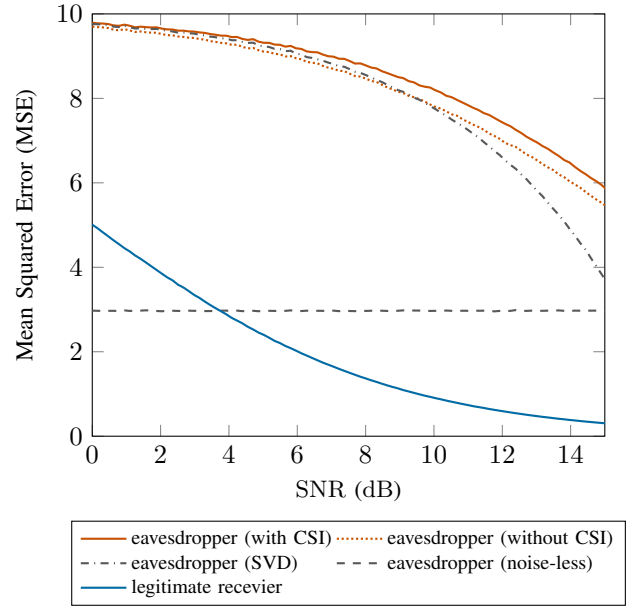
\begin{figure}[!htbp]
    \centering
    \input{mse_over_snr_dB}
    \caption{MSE at the legitimate receiver and at the eavesdropper over the received SNR at the legitimate receiver. The MSE at the eavesdropper is compared between different noise design methods.}
    \label{fig:mse_simulation}
\end{figure}

In this section, we present Monte Carlo simulation results on the MSE at the legitimate receiver and at the eavesdropper, respectively, over random realizations of the channel coefficients $\hbf$ and $\gbf$. The results are presented in \cref{fig:mse_simulation} and plotted over a given receive SNR at the legitimate receiver.
For each SNR, the results are averaged over \num{1e6} simulation runs. Note that, according to \eqref{eq:snr}, the SNR is proportional to the squared signal scaling factor, and the range of achievable SNR is limited by the smallest channel coefficient $h_1$. In our simulations, we choose the value $h_1$ at the standard deviation of a Rayleigh distribution with unit scale, i.e. $h_1=\left( \sqrt{\pi} - \sqrt{4-\pi} \right) / \sqrt{2}$. We draw the remaining channel coefficients $h_2,\dots,h_M$ and $g_1,\dots,g_M$ from a Rayleigh distribution with unit scale ($\sigma_g=\sigma_h=1$), but only consider cases in which $h_2,\dots,h_M \geq h_1$. We furthermore, consider the worst-case in which the eavesdropper's noise variance is $\sigma_z^2=0$, and set $\sigma_y^2=\num{0.1}$. For the sake of simplicity, we consider the case $\gradcov = \frac{1}{k} \Id$ and $k=1$. The simulations were run for $M=10$ users with power constraint $P=\num{1}$. We also compare our solution to the artificial-noise-free case and naive SVD approach. In the artificial-noise-free case, $\Abf=\zerobf$ is used. In the naive SVD approach, the precoding matrix is generated based on the SVD of the channel vector $\hbf$, which is an alternative basis for the channel null space. The obtained basis is then scaled by a common scalar in order to satisfy the power constraint.

The results show that the proposed secure over-the-air computation scheme is effective in the sense that it drastically increases the MSE at the eavesdropper compared to the legitimate receiver's MSE. Over a wide range of SNR values, it achieves MSE-security with a high security level, while the computation result is MSE-approximate with a good approximation level. While the figure shows the average case, the achievable security and approximation level can be higher or lower, depending on the instantaneous channel coefficients. The results also show that the MSE-security level can be increased by taking into account the eavesdropper's CSI in the system design. Furthermore, it can be observed that the proposed precoding design improves the security level over the artificial-noise-free case and the naive SVD approach. Since we set $\sigma_z^2=0$, the error at the eavesdropper in the artificial-noise-free case is purely caused by the channel mismatch, which is constant over the SNR.

%% file: mse_over_snr_dB.tex
\definecolor{myblue}{RGB}{0,107,164}
\definecolor{myred}{RGB}{200,82,0}
\definecolor{mygray}{RGB}{89,89,89}
 
\begin{tikzpicture}

\begin{axis}[%
width=0.94\linewidth,
xmin=0,
xmax=15,
xlabel style={font=\small},
xlabel={$\snr$ (\si{\decibel})},
ymin=0,
ymax=10,
ylabel style={font=\small},
ylabel={Mean Squared Error (MSE)},
legend columns=2,
legend style={at={(0.5,-0.2)},anchor=north,legend cell align=left, align=left, nodes={scale=0.75, transform shape}}
]

\addplot [color=myred,thick]
  table[row sep=crcr]{%
0	9.79185597915559\\
0.156935406450202	9.77622727049028\\
0.313870812900403	9.77203623767292\\
0.470806219350604	9.75890050886651\\
0.627741625800806	9.72727530117074\\
0.784677032251007	9.71804903489366\\
0.941612438701209	9.74542086853483\\
1.09854784515141	9.71928499903512\\
1.25548325160161	9.68249374067232\\
1.41241865805181	9.69781594989607\\
1.56935406450201	9.6865025610644\\
1.72628947095222	9.67919585620001\\
1.88322487740242	9.68168857603876\\
2.04016028385262	9.65118601707578\\
2.19709569030282	9.63320295260129\\
2.35403109675302	9.6300310196972\\
2.51096650320322	9.60360684214098\\
2.66790190965343	9.60359671725323\\
2.82483731610363	9.58615535858335\\
2.98177272255383	9.58130044354871\\
3.13870812900403	9.56214614722889\\
3.29564353545423	9.54997659154199\\
3.45257894190443	9.52379490636089\\
3.60951434835464	9.50659815757698\\
3.76644975480484	9.48531566949372\\
3.92338516125504	9.48442830765612\\
4.08032056770524	9.45144121315485\\
4.23725597415544	9.45449834891478\\
4.39419138060564	9.40726601755743\\
4.55112678705584	9.38439791550424\\
4.70806219350605	9.40198785956846\\
4.86499759995625	9.34289739438922\\
5.02193300640645	9.3157586820569\\
5.17886841285665	9.31487577052915\\
5.33580381930685	9.27367124314172\\
5.49273922575705	9.26910966204981\\
5.64967463220726	9.22494872801984\\
5.80661003865746	9.24221523128963\\
5.96354544510766	9.18622876451156\\
6.12048085155786	9.1704361345988\\
6.27741625800806	9.11127275293605\\
6.43435166445826	9.09575235566072\\
6.59128707090846	9.09026316130551\\
6.74822247735866	9.03715499378796\\
6.90515788380887	8.99825338744027\\
7.06209329025907	8.9870328758834\\
7.21902869670927	8.93882859001761\\
7.37596410315947	8.93568157989538\\
7.53289950960967	8.88435451519083\\
7.68983491605988	8.82465983147134\\
7.84677032251008	8.82811790090959\\
8.00370572896028	8.78115113299354\\
8.16064113541048	8.7301902265011\\
8.31757654186068	8.70935491047965\\
8.47451194831088	8.6442548144001\\
8.63144735476108	8.59947056270971\\
8.78838276121129	8.55204422336157\\
8.94531816766149	8.50500026827946\\
9.10225357411169	8.47924933177998\\
9.25918898056189	8.40856273761628\\
9.41612438701209	8.37840407247428\\
9.57305979346229	8.3078708223053\\
9.72999519991249	8.27592515065898\\
9.8869306063627	8.25669546600205\\
10.0438660128129	8.19129129571176\\
10.2008014192631	8.1411608277005\\
10.3577368257133	8.06247172060599\\
10.5146722321635	8.02847227338476\\
10.6716076386137	7.96510076975866\\
10.8285430450639	7.90401762405477\\
10.9854784515141	7.84183895413235\\
11.1424138579643	7.77934242076003\\
11.2993492644145	7.72523739968329\\
11.4562846708647	7.6523772720147\\
11.6132200773149	7.60750011864704\\
11.7701554837651	7.53557222323377\\
11.9270908902153	7.46391132443519\\
12.0840262966655	7.39697615993791\\
12.2409617031157	7.33858154485779\\
12.3978971095659	7.25366916680106\\
12.5548325160161	7.18062373975164\\
12.7117679224663	7.09008341773482\\
12.8687033289165	7.02005777205073\\
13.0256387353667	6.95250086990112\\
13.1825741418169	6.86869868707586\\
13.3395095482671	6.80349965307899\\
13.4964449547173	6.71549475214103\\
13.6533803611675	6.63656618722671\\
13.8103157676177	6.54048802935727\\
13.9672511740679	6.48118690324825\\
14.1241865805181	6.37671193117801\\
14.2811219869683	6.28435921287122\\
14.4380573934185	6.20538942841748\\
14.5949927998687	6.11386231221289\\
14.7519282063189	6.03827254910566\\
14.9088636127691	5.94775514187146\\
15.0657990192193	5.82912445757829\\
15.2227344256695	5.74568806344541\\
15.3796698321198	5.66011548803575\\
15.53660523857	5.54850675710702\\
};
\addlegendentry{eavesdropper (with CSI)}

\addplot [color=myred,densely dotted,thick]
  table[row sep=crcr]{%
0	9.69965714493284\\
0.156935406450202	9.68783267499974\\
0.313870812900403	9.68055375824633\\
0.470806219350604	9.66473113790541\\
0.627741625800806	9.62452053992767\\
0.784677032251007	9.61072666820515\\
0.941612438701209	9.63962259963307\\
1.09854784515141	9.60836477347851\\
1.25548325160161	9.57226002918601\\
1.41241865805181	9.58366681317437\\
1.56935406450201	9.56521706525363\\
1.72628947095222	9.55576530354566\\
1.88322487740242	9.5550619033557\\
2.04016028385262	9.51920375175493\\
2.19709569030282	9.49921753959007\\
2.35403109675302	9.49209827784208\\
2.51096650320322	9.46570079635057\\
2.66790190965343	9.45500778966206\\
2.82483731610363	9.43650019821843\\
2.98177272255383	9.43102056058377\\
3.13870812900403	9.40603344626773\\
3.29564353545423	9.3901687246878\\
3.45257894190443	9.35545045246274\\
3.60951434835464	9.33731922761024\\
3.76644975480484	9.31372320420753\\
3.92338516125504	9.30338316758917\\
4.08032056770524	9.26733059820492\\
4.23725597415544	9.26942482218931\\
4.39419138060564	9.22016283825791\\
4.55112678705584	9.18335149862852\\
4.70806219350605	9.19462488626356\\
4.86499759995625	9.13949714539854\\
5.02193300640645	9.10659227487804\\
5.17886841285665	9.09580385077196\\
5.33580381930685	9.05485626212496\\
5.49273922575705	9.04356849111472\\
5.64967463220726	8.99234156244211\\
5.80661003865746	9.00049266345136\\
5.96354544510766	8.95026412374158\\
6.12048085155786	8.92623909197351\\
6.27741625800806	8.86361904325785\\
6.43435166445826	8.84423799831107\\
6.59128707090846	8.82805102534638\\
6.74822247735866	8.7693684513031\\
6.90515788380887	8.72691456188572\\
7.06209329025907	8.70034323273702\\
7.21902869670927	8.65413573452955\\
7.37596410315947	8.64441297170488\\
7.53289950960967	8.58502116313028\\
7.68983491605988	8.52863600723272\\
7.84677032251008	8.51535480263331\\
8.00370572896028	8.46854520449064\\
8.16064113541048	8.41316210047088\\
8.31757654186068	8.38954221257406\\
8.47451194831088	8.32250942697177\\
8.63144735476108	8.26121093863537\\
8.78838276121129	8.21017474833027\\
8.94531816766149	8.15779364703534\\
9.10225357411169	8.12572921576753\\
9.25918898056189	8.05995065175161\\
9.41612438701209	8.02708687975238\\
9.57305979346229	7.98217131792059\\
9.72999519991249	7.92597201992957\\
9.8869306063627	7.8546329601903\\
10.0438660128129	7.79696721104859\\
10.2008014192631	7.75644885322263\\
10.3577368257133	7.69403542673308\\
10.5146722321635	7.6241402266352\\
10.6716076386137	7.59038810978909\\
10.8285430450639	7.51046859250761\\
10.9854784515141	7.4454208712066\\
11.1424138579643	7.37981140840156\\
11.2993492644145	7.32489332646525\\
11.4562846708647	7.25659676408397\\
11.6132200773149	7.16721907401257\\
11.7701554837651	7.11032722323152\\
11.9270908902153	7.0400976148173\\
12.0840262966655	6.94575964460535\\
12.2409617031157	6.89223283648326\\
12.3978971095659	6.84935385705687\\
12.5548325160161	6.7596890822121\\
12.7117679224663	6.66930740751702\\
12.8687033289165	6.61171413478041\\
13.0256387353667	6.52962961003655\\
13.1825741418169	6.44235132733968\\
13.3395095482671	6.36469001866548\\
13.4964449547173	6.29991455818251\\
13.6533803611675	6.1965518286302\\
13.8103157676177	6.13435079787786\\
13.9672511740679	6.03759326959397\\
14.1241865805181	5.96143763438887\\
14.2811219869683	5.88560793167932\\
14.4380573934185	5.78475049933373\\
14.5949927998687	5.69324049074711\\
14.7519282063189	5.60526266941626\\
14.9088636127691	5.52631965041299\\
15.0657990192193	5.42859139842395\\
15.2227344256695	5.33791159633629\\
15.3796698321198	5.26167706194622\\
15.53660523857	5.15154545132789\\
};
\addlegendentry{eavesdropper (without CSI)}

\addplot [dashdotted,thick,color=mygray]
  table[row sep=crcr]{%
0	9.76178339597486\\
0.156935406450202	9.75519401104116\\
0.313870812900403	9.74533691361247\\
0.470806219350604	9.73633866521505\\
0.627741625800806	9.69802314313268\\
0.784677032251007	9.68689883788667\\
0.941612438701209	9.71911634179876\\
1.09854784515141	9.68357460059012\\
1.25548325160161	9.65391603735473\\
1.41241865805181	9.66315797094437\\
1.56935406450201	9.6422555850574\\
1.72628947095222	9.64115277129635\\
1.88322487740242	9.64525198678456\\
2.04016028385262	9.61746437680834\\
2.19709569030282	9.6018943637208\\
2.35403109675302	9.5832686105328\\
2.51096650320322	9.55687052342619\\
2.66790190965343	9.55909448164492\\
2.82483731610363	9.54171286125868\\
2.98177272255383	9.53417503980446\\
3.13870812900403	9.50331197783595\\
3.29564353545423	9.49344334049103\\
3.45257894190443	9.46362253903312\\
3.60951434835464	9.44844320339726\\
3.76644975480484	9.42173613000507\\
3.92338516125504	9.4176300738904\\
4.08032056770524	9.37716318117004\\
4.23725597415544	9.37662820988572\\
4.39419138060564	9.33717742938367\\
4.55112678705584	9.31422287574001\\
4.70806219350605	9.31868264575826\\
4.86499759995625	9.25032684683787\\
5.02193300640645	9.23298072471599\\
5.17886841285665	9.2167750783716\\
5.33580381930685	9.17262225399875\\
5.49273922575705	9.16677285238231\\
5.64967463220726	9.1175102853138\\
5.80661003865746	9.12806918419447\\
5.96354544510766	9.05839314655537\\
6.12048085155786	9.03992151368547\\
6.27741625800806	8.97678322797796\\
6.43435166445826	8.95740904243106\\
6.59128707090846	8.92325347259056\\
6.74822247735866	8.89217586947787\\
6.90515788380887	8.84705217383807\\
7.06209329025907	8.81337749309479\\
7.21902869670927	8.76029340580978\\
7.37596410315947	8.74815011592161\\
7.53289950960967	8.6916782171817\\
7.68983491605988	8.63065837798582\\
7.84677032251008	8.60301270713573\\
8.00370572896028	8.55347092006513\\
8.16064113541048	8.49650899064678\\
8.31757654186068	8.4367191047343\\
8.47451194831088	8.38753017480819\\
8.63144735476108	8.32301361314613\\
8.78838276121129	8.25819050907514\\
8.94531816766149	8.19610421431343\\
9.10225357411169	8.15439046937992\\
9.25918898056189	8.09082483598285\\
9.41612438701209	8.01609372576265\\
9.57305979346229	7.96819852293274\\
9.72999519991249	7.89651845299811\\
9.8869306063627	7.8202640968244\\
10.0438660128129	7.74581118403638\\
10.2008014192631	7.66072196603055\\
10.3577368257133	7.59214336855129\\
10.5146722321635	7.49967958937735\\
10.6716076386137	7.42287750316\\
10.8285430450639	7.34711083875712\\
10.9854784515141	7.25883918105692\\
11.1424138579643	7.16385914513646\\
11.2993492644145	7.07959763344876\\
11.4562846708647	6.967075354655\\
11.6132200773149	6.86365981662034\\
11.7701554837651	6.75549984640689\\
11.9270908902153	6.65573425463337\\
12.0840262966655	6.53205826246898\\
12.2409617031157	6.42513144688498\\
12.3978971095659	6.32877985986456\\
12.5548325160161	6.20154001331115\\
12.7117679224663	6.08344157526559\\
12.8687033289165	5.94066630435313\\
13.0256387353667	5.80616879069636\\
13.1825741418169	5.67847818158374\\
13.3395095482671	5.52642458906263\\
13.4964449547173	5.38040538842721\\
13.6533803611675	5.24171932968273\\
13.8103157676177	5.08617291633713\\
13.9672511740679	4.91216716306377\\
14.1241865805181	4.75384683563625\\
14.2811219869683	4.5743157552975\\
14.4380573934185	4.4172448537894\\
14.5949927998687	4.21654595106159\\
14.7519282063189	4.02682745966453\\
14.9088636127691	3.8320103831122\\
15.0657990192193	3.63627288744593\\
15.2227344256695	3.41844546268364\\
15.3796698321198	3.19112336362958\\
15.53660523857	2.97613996403193\\
};
\addlegendentry{eavesdropper (SVD)}

\addplot [thick,dashed,color=mygray]
  table[row sep=crcr]{%
0	2.97406028428287\\
0.156935406450202	2.96927206126454\\
0.313870812900403	2.97034623617064\\
0.470806219350604	2.96926332989791\\
0.627741625800806	2.96866434905319\\
0.784677032251007	2.96764691948052\\
0.941612438701209	2.97971422167756\\
1.09854784515141	2.96613149114721\\
1.25548325160161	2.97593644391522\\
1.41241865805181	2.97422092434743\\
1.56935406450201	2.98102820473024\\
1.72628947095222	2.97479340935331\\
1.88322487740242	2.96552433798952\\
2.04016028385262	2.95865851646787\\
2.19709569030282	2.9748855389724\\
2.35403109675302	2.9705744317331\\
2.51096650320322	2.96714076246758\\
2.66790190965343	2.96286910362114\\
2.82483731610363	2.97261115410115\\
2.98177272255383	2.96903847457702\\
3.13870812900403	2.97206964542062\\
3.29564353545423	2.96346702350901\\
3.45257894190443	2.96376988469213\\
3.60951434835464	2.96641520110303\\
3.76644975480484	2.97274901120518\\
3.92338516125504	2.97362425256115\\
4.08032056770524	2.96443985095759\\
4.23725597415544	2.97431626151943\\
4.39419138060564	2.96129801240827\\
4.55112678705584	2.96535729638597\\
4.70806219350605	2.96978970605989\\
4.86499759995625	2.95988018079349\\
5.02193300640645	2.96625826447469\\
5.17886841285665	2.96955049846856\\
5.33580381930685	2.96676528812722\\
5.49273922575705	2.96591109748951\\
5.64967463220726	2.96857277006258\\
5.80661003865746	2.97972162139226\\
5.96354544510766	2.96297446940016\\
6.12048085155786	2.96437095858041\\
6.27741625800806	2.97250938423432\\
6.43435166445826	2.96618527371208\\
6.59128707090846	2.96969723204879\\
6.74822247735866	2.9686033164867\\
6.90515788380887	2.9705848141816\\
7.06209329025907	2.97356595425516\\
7.21902869670927	2.96143885562779\\
7.37596410315947	2.97506880290291\\
7.53289950960967	2.96493432667468\\
7.68983491605988	2.96140969924233\\
7.84677032251008	2.97306846742962\\
8.00370572896028	2.9666377429901\\
8.16064113541048	2.97085158797611\\
8.31757654186068	2.96568830619238\\
8.47451194831088	2.97193146516937\\
8.63144735476108	2.96638499553424\\
8.78838276121129	2.96242429860091\\
8.94531816766149	2.9707263224353\\
9.10225357411169	2.97988941929345\\
9.25918898056189	2.97138916024004\\
9.41612438701209	2.97678036731901\\
9.57305979346229	2.97477037685063\\
9.72999519991249	2.9764969171701\\
9.8869306063627	2.96591833700348\\
10.0438660128129	2.96928520056991\\
10.2008014192631	2.96498741317755\\
10.3577368257133	2.97585598124878\\
10.5146722321635	2.96949997948902\\
10.6716076386137	2.96630798387501\\
10.8285430450639	2.97124043809328\\
10.9854784515141	2.97034011766409\\
11.1424138579643	2.97486502085464\\
11.2993492644145	2.97100319543055\\
11.4562846708647	2.97258514540318\\
11.6132200773149	2.96857641591058\\
11.7701554837651	2.9564548546688\\
11.9270908902153	2.96705394890128\\
12.0840262966655	2.96260877631866\\
12.2409617031157	2.96877126062195\\
12.3978971095659	2.98723828282113\\
12.5548325160161	2.97105154922811\\
12.7117679224663	2.96768940804173\\
12.8687033289165	2.97363644470108\\
13.0256387353667	2.96821585722302\\
13.1825741418169	2.96983869045638\\
13.3395095482671	2.97297235534304\\
13.4964449547173	2.97482744994546\\
13.6533803611675	2.97947127770317\\
13.8103157676177	2.96878295960376\\
13.9672511740679	2.96577775812974\\
14.1241865805181	2.97381676858021\\
14.2811219869683	2.97195362944501\\
14.4380573934185	2.96957253040882\\
14.5949927998687	2.97350487383269\\
14.7519282063189	2.97398258661595\\
14.9088636127691	2.96503078670196\\
15.0657990192193	2.96887020065868\\
15.2227344256695	2.96985360492987\\
15.3796698321198	2.96483469552716\\
15.53660523857	2.97411881229686\\
};
\addlegendentry{eavesdropper (noise-less)}

\addplot [color=myblue,thick]
  table[row sep=crcr]{%
0	5.00765404998283\\
0.156935406450202	4.92263813024302\\
0.313870812900403	4.82547401636212\\
0.470806219350604	4.73172361602447\\
0.627741625800806	4.63577096358391\\
0.784677032251007	4.54590771630714\\
0.941612438701209	4.45274599051141\\
1.09854784515141	4.37352605937216\\
1.25548325160161	4.27959163431244\\
1.41241865805181	4.20143642928986\\
1.56935406450201	4.11463129341528\\
1.72628947095222	4.02442975636151\\
1.88322487740242	3.93653174737829\\
2.04016028385262	3.84640978581939\\
2.19709569030282	3.75952466007833\\
2.35403109675302	3.68888214769858\\
2.51096650320322	3.58717603859202\\
2.66790190965343	3.52127147061574\\
2.82483731610363	3.43589443406439\\
2.98177272255383	3.34533904526247\\
3.13870812900403	3.27426909243608\\
3.29564353545423	3.18749623763546\\
3.45257894190443	3.1068224052986\\
3.60951434835464	3.03139453345165\\
3.76644975480484	2.95890170399811\\
3.92338516125504	2.8860198142391\\
4.08032056770524	2.80235911817518\\
4.23725597415544	2.73811704979725\\
4.39419138060564	2.67303282661127\\
4.55112678705584	2.59188483346357\\
4.70806219350605	2.52781819959859\\
4.86499759995625	2.4548467354802\\
5.02193300640645	2.39844267662285\\
5.17886841285665	2.33141368852121\\
5.33580381930685	2.2576963053643\\
5.49273922575705	2.20521947205735\\
5.64967463220726	2.13754109099912\\
5.80661003865746	2.08518697220983\\
5.96354544510766	2.02565828004332\\
6.12048085155786	1.96372596930072\\
6.27741625800806	1.90939739841973\\
6.43435166445826	1.85333969333074\\
6.59128707090846	1.79838612442218\\
6.74822247735866	1.74241580622197\\
6.90515788380887	1.69475305429635\\
7.06209329025907	1.64557368256162\\
7.21902869670927	1.59547546800537\\
7.37596410315947	1.54747732810039\\
7.53289950960967	1.5006505991354\\
7.68983491605988	1.45465990995202\\
7.84677032251008	1.40857347020772\\
8.00370572896028	1.36816501777763\\
8.16064113541048	1.32446517008548\\
8.31757654186068	1.28600494208559\\
8.47451194831088	1.24374889321434\\
8.63144735476108	1.20414907708265\\
8.78838276121129	1.16647524128185\\
8.94531816766149	1.13321461925539\\
9.10225357411169	1.09248360077075\\
9.25918898056189	1.05989515302519\\
9.41612438701209	1.0260064048798\\
9.57305979346229	0.992475651573072\\
9.72999519991249	0.959923332483172\\
9.8869306063627	0.93233975385415\\
10.0438660128129	0.901436929052179\\
10.2008014192631	0.874105626522475\\
10.3577368257133	0.843982986823338\\
10.5146722321635	0.816966493543263\\
10.6716076386137	0.790177875020139\\
10.8285430450639	0.763679506257993\\
10.9854784515141	0.739517132471828\\
11.1424138579643	0.711702733108111\\
11.2993492644145	0.690727040639552\\
11.4562846708647	0.669740116966716\\
11.6132200773149	0.644788151964044\\
11.7701554837651	0.623854289212745\\
11.9270908902153	0.603008674363379\\
12.0840262966655	0.583307898943467\\
12.2409617031157	0.562951317956416\\
12.3978971095659	0.544286043715859\\
12.5548325160161	0.526160671747114\\
12.7117679224663	0.507976558185071\\
12.8687033289165	0.490622984381167\\
13.0256387353667	0.474421278064864\\
13.1825741418169	0.457692511813301\\
13.3395095482671	0.443713533877259\\
13.4964449547173	0.427682011077687\\
13.6533803611675	0.412861758323116\\
13.8103157676177	0.400231175148921\\
13.9672511740679	0.385178688035686\\
14.1241865805181	0.373003663795903\\
14.2811219869683	0.359271712572452\\
14.4380573934185	0.34697574383455\\
14.5949927998687	0.335872421868068\\
14.7519282063189	0.323881090646772\\
14.9088636127691	0.312907470173902\\
15.0657990192193	0.302460011871301\\
15.2227344256695	0.291773435241986\\
15.3796698321198	0.281222614853011\\
15.53660523857	0.271773912809894\\
};
\addlegendentry{legitimate recevier}

\end{axis}

\end{tikzpicture}%

%% file: appendices.tex
\section{Proof of \cref{thm:DvsS}}
\label{apx:DvsS}
The estimator that minimizes the MSE is given by the expected value of the posterior distribution~\cite[p. 143]{poorIntroductionSignalDetection1998}. Furthermore, the MSE of this estimator is given by the trace of the conditional covariance $\Trace\left( \Cov\left[ \sbf \mid \ybf \right] \right)$~\cite[p. 155]{poorIntroductionSignalDetection1998}.

Note that $\sbf$ and $\ybf$ are jointly Gaussian random vectors. For this special case, the MSE of the optimal estimator\footnote{The optimal estimator for the legitimate receiver is $\hat{\sbf}_y = \Exp\left[ \sbf \mid \ybf \right] =  \Cov\left[\sbf,\ybf\right] \Cov\left[\ybf\right]^{-1} \ybf$.}
at the legitimate receiver is given by $\Trace\left( \Cov\left[ \sbf \right] - \Cov\left[ \sbf,\ybf \right] \Cov\left[ \ybf \right]^{-1} \Cov\left[ \ybf,\sbf \right] \right)$.
We first compute
\begin{align*}
    \Cov\left[\sbf\right] &= M \gradcov, \\
    \Cov\left[\sbf,\ybf\right] &= \Cov\left[\ybf,\sbf\right] = c M \gradcov, \\
    \Cov\left[\ybf\right] &= c^2 M \gradcov + \sigma_y^2 \Id,
\end{align*}
and thus, obtain \eqref{eq:D-approximate} as
\begin{align*}
    D &= \Trace\left( M \gradcov - c^2 M^2 \gradcov \left( c^2 M \gradcov + \sigma_y^2 \Id \right)^{-1} \gradcov \right) \\
    &= \Trace\left( M \gradcov \right) - \Trace\left( c^2 M^2 \gradcov \left( c^2 M \gradcov + \sigma_y^2 \Id \right)^{-1} \gradcov \right) \\
    &= M - c^2 M^2 \Trace\left( \gradcov \left( c^2 M \gradcov + \sigma_y^2 \Id \right)^{-1} \gradcov \right).
\end{align*}

By the same arguments, we note that for every estimator $d_z\left(\zbf\right)$ at the eavesdropper it holds
$\Exp\left[ \left\| d\left(\zbf\right) - \sbf \right\|^2 \right] \geq \Exp\left[ \left\| \Exp\left[ \sbf \mid \zbf \right] - \sbf \right\|^2 \right]$. First, let $\vbf_i^\transpose$ be the $i$-th row of $\Vbf$ and compute the covariance matrix $\Cov\left[ \Vbf \Abf \gbf \right]$ by
\begin{align*}
    &\Cov\left[ \Vbf \Abf \gbf \right]_{ij} = \Exp\left[ \Vbf \left( \Abf \gbf \gbf^\transpose \Abf^\transpose \right) \Vbf^\transpose \right]_{ij} \\
    =& \Exp\left[ \vbf_i^\transpose \left( \Abf \gbf \gbf^\transpose \Abf^\transpose \right) \vbf_j \right]\\
    =&
    \begin{cases}
    \Exp\left[ \vbf_i^\transpose \right] \left( \Abf \gbf \gbf^\transpose \Abf^\transpose \right) \Exp\left[ \vbf_j \right] = 0 & \text{if } i \neq j\\
    \Exp\left[ \vbf_i^\transpose \left( \Abf \gbf \gbf^\transpose \Abf^\transpose \right) \vbf_i \right] = \left\| \Abf \gbf \right\|^2 & \text{if } i = j
    \end{cases}
    .
\end{align*}
From this, we can compute the covariance matrices
\begin{align*}
    \Cov\left[ \zbf \right] &= \Cov\left[ c \sum_{m=1}^{M} \frac{g_m}{h_m} \grad_m \right] + \Cov\left[ \Vbf \Abf \gbf \right] + \Cov\left[ \nbf_z \right]\\
    &= c^2 \left(\sum_{m=1}^{M}\frac{g_m^2}{h_m^2}\right) \gradcov + \left( \left\| \Abf \gbf \right\|^2 + \sigma_z^2 \right) \Id,
\end{align*}
and
\begin{align*}
    &\Cov\left[ \sbf, \zbf \right] \\
    =& \Exp\left[ \left( c \sum_{m=1}^{M} \frac{g_m}{h_m} \grad_m + \Vbf \Abf \gbf + \nbf_z \right) \left( \sum_{m=1}^{M} \grad_m \right)^\transpose \right] \\
    \overset{a)}{=}& c \Exp\left[ \sum_{l=1}^{M}\sum_{m=1}^{M} \frac{g_m}{h_m} \grad_m \grad_l^\transpose \right] \\
    \overset{b)}{=}& c \sum_{m=1}^{M} \frac{g_m}{h_m} \Exp\left[ \grad_m \grad_m^\transpose \right] = c \left(\sum_{m=1}^{M} \frac{g_m}{h_m}\right) \gradcov,
\end{align*}
where $a)$ and $b)$ follow from the independence between $\grad_m$, $\Vbf$, and $\nbf_z$. 

We finally obtain
\begin{align*}
    S &= \Trace\left( \Cov\left[ \sbf \right] - \Cov\left[ \sbf,\ybf \right] \Cov\left[ \ybf \right]^{-1} \Cov\left[ \ybf,\sbf \right] \right) \\
    &= M - c^2 \left( \sum_{m=1}^{M} \frac{g_m}{h_m} \right)^2 \\
     & \Trace\left( \gradcov \left[ c^2 \left( \sum_{m=1}^{M} \frac{g_m^2}{h_m^2} \right) \gradcov + \left( \left\| \Abf \gbf \right\|^2 + \sigma_z^2 \right) \Id \right]^{-1} \gradcov \right).
\end{align*}

\section{Proof of \cref{thm:mse_requirement}}
\label{apx:mse_requirement}
From the MSE requirement $\mu$ and \eqref{eq:D-approximate} it follows
\begin{equation*}
M-\mu \leq c^2 M^2 \Trace\left( \left[ c^2 M \gradcov^{-1} + \sigma_y^2 \left( \gradcov^2 \right)^{-1} \right]^{-1} \right).
\end{equation*}
Let $\lambda_1 \geq \dots \geq \lambda_k$ be the eigenvalues of $\gradcov$. Then the eigenvalues of $\Pbf = c^2 M \gradcov^{-1}$ are $\frac{c^2 M}{\lambda_i}$ and the eigenvalues of $\Qbf = \sigma_y^2 \left( \gradcov^2 \right)^{-1}$ are $\frac{\sigma_y^2}{\lambda_i^2}$. If we denote the eigenvalues of $\Pbf + \Qbf$ by $\kappa_1 \geq \dots \geq \kappa_n$, we have
\begin{equation}
\label{eq:sum_bound}
M-\mu \leq c^2 M^2 \sum_{i=1}^{n} \frac{1}{\kappa_i} \leq c^2 M^2 n \max_{i=1,\dots,n} \frac{1}{\kappa_i} = c^2 M^2 \frac{n}{\kappa_n}
\end{equation}
We use that for real symmetric matrices $\Pbf$ and $\Qbf$, the smallest eigenvalue $\kappa_n$ of their sum is lower-bounded by the sum of the smallest eigenvalue of $\Pbf$ and $\Qbf$ respectively~\cite[Eq. 2]{fultonEigenvaluesInvariantFactors2000}. We thus, have
\begin{equation}
\label{eq:ev_bound}
\kappa_n \geq \frac{c^2M}{\lambda_1} + \frac{\sigma_y^2}{\lambda_1^2} = \frac{c^2 M \lambda_1 + \sigma_y^2}{\lambda_1^2},    
\end{equation}
and we can write
\begin{align}
& M-\mu &\leq& \frac{c^2 M^2 n \lambda_1^2}{c^2 M \lambda_1 + \sigma_y^2}
\label{eq:MSE_bound}
\\
\Leftrightarrow & c^2 &\geq& \frac{\sigma_y^2 (M-\mu)}{M^2 n \lambda_1^2 - M(M-\mu) \lambda_1}.
\label{eq:c_bound}
\end{align}
Finally, we conclude the proof by combining \eqref{eq:c_bound} with the upper bound from \eqref{eq:c} maximized over $\left\| \abf_m \right\|^2$.

%% file: main.bbl
\begin{thebibliography}{10}
\providecommand{\url}[1]{#1}
\csname url@samestyle\endcsname
\providecommand{\newblock}{\relax}
\providecommand{\bibinfo}[2]{#2}
\providecommand{\BIBentrySTDinterwordspacing}{\spaceskip=0pt\relax}
\providecommand{\BIBentryALTinterwordstretchfactor}{4}
\providecommand{\BIBentryALTinterwordspacing}{\spaceskip=\fontdimen2\font plus
\BIBentryALTinterwordstretchfactor\fontdimen3\font minus
  \fontdimen4\font\relax}
\providecommand{\BIBforeignlanguage}[2]{{%
\expandafter\ifx\csname l@#1\endcsname\relax
\typeout{** WARNING: IEEEtran.bst: No hyphenation pattern has been}%
\typeout{** loaded for the language `#1'. Using the pattern for}%
\typeout{** the default language instead.}%
\else
\language=\csname l@#1\endcsname
\fi
#2}}
\providecommand{\BIBdecl}{\relax}
\BIBdecl

\bibitem{khan6GWirelessSystems2020}
L.~U. Khan, I.~Yaqoob, M.~Imran, Z.~Han, and C.~S. Hong, ``{{6G Wireless
  Systems}}: {{A Vision}}, {{Architectural Elements}}, and {{Future
  Directions}},'' \emph{IEEE Access}, vol.~8, pp. 147\,029--147\,044, 2020.

\bibitem{mcmahanCommunicationEfficientLearningDeep2017}
B.~McMahan, E.~Moore, D.~Ramage, S.~Hampson, and B.~A. y~Arcas,
  ``Communication-{{Efficient Learning}} of {{Deep Networks}} from
  {{Decentralized Data}},'' in \emph{Proceedings of the 20th {{International
  Conference}} on {{Artificial Intelligence}} and {{Statistics}}}.\hskip 1em
  plus 0.5em minus 0.4em\relax {PMLR}, Apr. 2017, pp. 1273--1282.

\bibitem{nazerComputationMultipleAccessChannels2007}
B.~Nazer and M.~Gastpar, ``Computation {{Over Multiple-Access Channels}},''
  \emph{IEEE Transactions on Information Theory}, vol.~53, no.~10, pp.
  3498--3516, Oct. 2007.

\bibitem{goldenbaumHarnessingInterferenceAnalog2013}
M.~Goldenbaum, H.~Boche, and S.~Sta{\'n}czak, ``Harnessing {{Interference}} for
  {{Analog Function Computation}} in {{Wireless Sensor Networks}},'' \emph{IEEE
  Transactions on Signal Processing}, vol.~61, no.~20, pp. 4893--4906, Oct.
  2013.

\bibitem{abariOvertheairFunctionComputation2016}
\BIBentryALTinterwordspacing
O.~Abari, H.~Rahul, and D.~Katabi, ``Over-the-air {{Function Computation}} in
  {{Sensor Networks}},'' 2016. [Online]. Available:
  \url{https://arxiv.org/abs/1612.02307}
\BIBentrySTDinterwordspacing

\bibitem{goldenbaumRobustAnalogFunction2013}
M.~Goldenbaum and S.~Stanczak, ``Robust {{Analog Function Computation}} via
  {{Wireless Multiple-Access Channels}},'' \emph{IEEE Transactions on
  Communications}, vol.~61, no.~9, pp. 3863--3877, Sep. 2013.

\bibitem{zhuBroadbandAnalogAggregation2020}
G.~Zhu, Y.~Wang, and K.~Huang, ``Broadband {{Analog Aggregation}} for
  {{Low-Latency Federated Edge Learning}},'' \emph{IEEE Transactions on
  Wireless Communications}, vol.~19, no.~1, pp. 491--506, Jan. 2020.

\bibitem{yangFederatedLearningOvertheAir2020}
K.~Yang, T.~Jiang, Y.~Shi, and Z.~Ding, ``Federated {{Learning}} via
  {{Over-the-Air Computation}},'' \emph{IEEE Transactions on Wireless
  Communications}, vol.~19, no.~3, pp. 2022--2035, Mar. 2020.

\bibitem{seryOvertheAirFederatedLearning2021}
T.~Sery, N.~Shlezinger, K.~Cohen, and Y.~C. Eldar, ``Over-the-{{Air Federated
  Learning From Heterogeneous Data}},'' \emph{IEEE Transactions on Signal
  Processing}, vol.~69, pp. 3796--3811, 2021.

\bibitem{amiriFederatedLearningWireless2020}
M.~M. Amiri and D.~G{\"u}nd{\"u}z, ``Federated {{Learning Over Wireless Fading
  Channels}},'' \emph{IEEE Transactions on Wireless Communications}, vol.~19,
  no.~5, pp. 3546--3557, May 2020.

\bibitem{huSecureTransceiverDesign2022}
C.~Hu, Q.~Li, Q.~Zhang, and J.~Qin, ``Secure {{Transceiver Design}} and {{Power
  Control}} for {{Over-the-Air Computation Networks}},'' \emph{IEEE
  Communications Letters}, vol.~26, no.~7, pp. 1509--1513, Apr. 2022.

\bibitem{yanSecurePrivateOvertheAir2022}
\BIBentryALTinterwordspacing
N.~Yan, K.~Wang, K.~Zhi, C.~Pan, K.~K. Chai, and H.~V. Poor, ``Toward
  {{Secure}} and {{Private Over-the-Air Federated Learning}},'' 2022. [Online].
  Available: \url{https://arxiv.org/abs/2210.07669}
\BIBentrySTDinterwordspacing

\bibitem{freySecureOverTheAirComputation2021}
M.~Frey, I.~Bjelakovi{\'c}, and S.~Sta{\'n}czak, ``Towards {{Secure
  Over-The-Air Computation}},'' in \emph{2021 {{IEEE International Symposium}}
  on {{Information Theory}} ({{ISIT}})}, Jul. 2021, pp. 700--705.

\bibitem{goelGuaranteeingSecrecyUsing2008}
S.~Goel and R.~Negi, ``Guaranteeing {{Secrecy}} using {{Artificial Noise}},''
  \emph{IEEE Transactions on Wireless Communications}, vol.~7, no.~6, pp.
  2180--2189, Jun. 2008.

\bibitem{swindlehurstFixedSINRSolutions2009}
A.~L. Swindlehurst, ``Fixed {{SINR}} solutions for the {{MIMO}} wiretap
  channel,'' in \emph{2009 {{IEEE International Conference}} on {{Acoustics}},
  {{Speech}} and {{Signal Processing}}}, Apr. 2009, pp. 2437--2440.

\bibitem{linSecrecyRateGeneralized2013}
P.-H. Lin, S.-H. Lai, S.-C. Lin, and H.-J. Su, ``On {{Secrecy Rate}} of the
  {{Generalized Artificial-Noise Assisted Secure Beamforming}} for {{Wiretap
  Channels}},'' \emph{IEEE Journal on Selected Areas in Communications},
  vol.~31, no.~9, pp. 1728--1740, Sep. 2013.

\bibitem{liaoOvertheAirFederatedLearning2022}
J.~Liao, Z.~Chen, and E.~G. Larsson, ``Over-the-{{Air Federated Learning}} with
  {{Privacy Protection}} via {{Correlated Additive Perturbations}},'' in
  \emph{2022 58th {{Annual Allerton Conference}} on {{Communication}},
  {{Control}}, and {{Computing}} ({{Allerton}})}, Sep. 2022, pp. 1--8.

\bibitem{buckApproximateComplexityFunctional1979}
R.~C. Buck, ``Approximate complexity and functional representation,''
  \emph{Journal of Mathematical Analysis and Applications}, vol.~70, no.~1, pp.
  280--298, Jul. 1979.

\bibitem{goldenbaumNomographicFunctionsEfficient2015}
M.~Goldenbaum, H.~Boche, and S.~Sta{\'n}czak, ``Nomographic {{Functions}}:
  {{Efficient Computation}} in {{Clustered Gaussian Sensor Networks}},''
  \emph{IEEE Transactions on Wireless Communications}, vol.~14, no.~4, pp.
  2093--2105, Apr. 2015.

\bibitem{gastparUncodedTransmissionExactly2008}
M.~Gastpar, ``Uncoded {{Transmission Is Exactly Optimal}} for a {{Simple
  Gaussian}} ``{{Sensor}}'' {{Network}},'' \emph{IEEE Transactions on
  Information Theory}, vol.~54, no.~11, pp. 5247--5251, Nov. 2008.

\bibitem{poorIntroductionSignalDetection1998}
H.~V. Poor, \emph{An {{Introduction}} to {{Signal Detection}} and
  {{Estimation}}}.\hskip 1em plus 0.5em minus 0.4em\relax {Springer Science \&
  Business Media}, Mar. 1998.

\bibitem{fultonEigenvaluesInvariantFactors2000}
W.~Fulton, ``Eigenvalues, invariant factors, highest weights, and {{Schubert}}
  calculus,'' \emph{Bulletin of the American Mathematical Society}, vol.~37,
  no.~3, pp. 209--249, Apr. 2000.

\end{thebibliography}
